\documentclass[aps,letterpaper,10pt,twocolumn,showpacs,superscriptaddress]{revtex4-1}

\usepackage{graphicx}
\usepackage{dcolumn}
\usepackage{bm}

\usepackage[utf8]{inputenc}
\usepackage[T1]{fontenc}
\usepackage{mathptmx}
\usepackage{braket}
\usepackage{ulem}
\usepackage{siunitx}
\usepackage{amsfonts,amssymb,amsmath,amsthm,graphicx}
\usepackage[colorlinks=true, allcolors=blue]{hyperref}
\usepackage{xcolor}
\DeclareMathAlphabet{\mathcal}{OMS}{cmsy}{m}{n}

\begin{document}

\title[Resource-effective Quantum Key Distribution: a field-trial in Padua city center]{Resource-effective Quantum Key Distribution: a field-trial in Padua city center}

\author{Marco Avesani}
\thanks{These authors contributed equally to this work.}
\affiliation{Dipartimento di Ingegneria dell'Informazione, Universit\`a degli Studi di Padova, via Gradenigo 6B, 35131 Padova, Italy} 

\author{Luca Calderaro}
\thanks{These authors contributed equally to this work.}
\affiliation{Dipartimento di Ingegneria dell'Informazione, Universit\`a degli Studi di Padova, via Gradenigo 6B, 35131 Padova, Italy} 

\author{Giulio Foletto}
\thanks{These authors contributed equally to this work.}
\affiliation{Dipartimento di Ingegneria dell'Informazione, Universit\`a degli Studi di Padova, via Gradenigo 6B, 35131 Padova, Italy}

\author{Costantino Agnesi}
\affiliation{Dipartimento di Ingegneria dell'Informazione, Universit\`a degli Studi di Padova, via Gradenigo 6B, 35131 Padova, Italy}
\author{Francesco Picciariello}
\affiliation{Dipartimento di Ingegneria dell'Informazione, Universit\`a degli Studi di Padova, via Gradenigo 6B, 35131 Padova, Italy}
\author{Francesco Santagiustina}
\affiliation{Dipartimento di Ingegneria dell'Informazione, Universit\`a degli Studi di Padova, via Gradenigo 6B, 35131 Padova, Italy}

\author{Alessia Scriminich}
\affiliation{Dipartimento di Ingegneria dell'Informazione, Universit\`a degli Studi di Padova, via Gradenigo 6B, 35131 Padova, Italy}
\author{Andrea Stanco}
\affiliation{Dipartimento di Ingegneria dell'Informazione, Universit\`a degli Studi di Padova, via Gradenigo 6B, 35131 Padova, Italy}
\author{Francesco Vedovato}
\affiliation{Dipartimento di Ingegneria dell'Informazione, Universit\`a degli Studi di Padova, via Gradenigo 6B, 35131 Padova, Italy}
\author{Mujtaba Zahidy}
\affiliation{Dipartimento di Ingegneria dell'Informazione, Universit\`a degli Studi di Padova, via Gradenigo 6B, 35131 Padova, Italy}

\author{Giuseppe Vallone}
\affiliation{Dipartimento di Ingegneria dell'Informazione, Universit\`a degli Studi di Padova, via Gradenigo 6B, 35131 Padova, Italy}
\affiliation{Dipartimento di Fisica e Astronomia, Universit\`a degli Studi di Padova, via Marzolo 8, 35131 Padova, Italy}
\affiliation{Padua Quantum Technology Research Center, Universit\`a degli Studi di Padova}

\author{Paolo Villoresi}
\email[E-mail:]{paolo.villoresi@dei.unipd.it}
\homepage{https://quantumfuture.dei.unipd.it}
\affiliation{Dipartimento di Ingegneria dell'Informazione, Universit\`a degli Studi di Padova, via Gradenigo 6B, 35131 Padova, Italy}
\affiliation{Padua Quantum Technology Research Center, Universit\`a degli Studi di Padova}

\date{\today}

\begin{abstract}
Field-trials are of key importance for novel technologies seeking commercialization and wide-spread adoption. This is certainly also the case for Quantum Key Distribution (QKD), which allows distant parties to distill a secret key with unconditional security. Typically, QKD demonstrations over urban infrastructures require complex stabilization and synchronization systems to maintain a low Quantum Bit Error (QBER) and high secret key rates over time. Here we present a field-trial which exploits a low-complexity self-stabilized hardware and a novel synchronization technique, to perform QKD over optical fibers deployed in the city center of Padua, Italy. In particular, two techniques recently introduced by our research group are evaluated in a real-world environment: the iPOGNAC polarization encoder was used for the preparation of the quantum states, while the temporal synchronization was performed using the Qubit4Sync algorithm. The results here presented demonstrate the validity and robustness of our 
resource-effective QKD system, that can be easily and rapidly installed in an existing telecommunication infrastructure, thus representing an important step towards mature, efficient and low-cost QKD systems. 
\end{abstract}

\maketitle

\section{Introduction}

Broad interest in quantum technologies has led to the recent announcement of several national and international dedicated initiatives~\cite{Riedel_2019,Raymer_2019}.
All these actions have the common goal of propelling quantum technologies from scientific novelties that exist only as table-top experiments in research facilities to commercially available devices that have significant impacts on our everyday lives.
An example of such quantum technologies is Quantum Key Distribution (QKD),  which allows distant parties to distill a perfectly-secret key and bound the shared information with any adversarial eavesdropper~\cite{Bennett2014_BB84, Gisin_review2002}.
Supported by a solid theoretical basis~\cite{Scarani2008} and ongoing technical innovations~\cite{Pirandola2019rev}, QKD is currently one of the most mature among quantum technologies, and has already performed the leap towards commercialization. 
In fact, the recent efforts of several research groups have focused on devising simpler protocols~\cite{Fung2006,Rusca2018} and developing more robust and economical experimental implementations~\cite{Duligall_2006,Sibson:17,Boaron2018,Xia:2019,Agnesi:20}, aiming at widespread adoption of QKD in  current telecommunication networks.  
However, efficient design, resource-effectiveness, and field-trials are necessary to bridge the gap between table-top experiments and commercially available devices~\cite{Stucki_2011, Wang:14,Bunandar2018, Dynes2019}
paving the way towards QKD standardization.  Indeed, these field-trials allow  the performance evaluation of the developed systems in real-world conditions that include an uncontrolled environment and quantum channel, as well as their suitability for exploitation in concrete applications.

Here we report a QKD field-trial performed in the metropolitan area of Padua, Italy. The field-trial exploited the optical fibre of the interdepartment network of the University of Padua which connects the Department of Mathematics with the ICT Center (\textit{Area Servizi Informatici e Telematici}, ASIT) of the university. The polarization of weak coherent pulses was exploited to implement a simplified three-state and one-decoy BB84 QKD protocol~\cite{Grunenfelder2018}. Polarization encoding was performed using the iPOGNAC polarization encoder~\cite{Avesani:2020} (patent pending~\footnote{The iPOGNAC is object of the Italian Patent Application No. 102019000019373 filed on 21.10.2019 as well as of the International Patent Application no. PCT/EP2020/079471 filed on 20.10.2020.}), while temporal synchronization between the transmitter and the receiver was realized by exploiting the Qubit4Sync method~\cite{Calderaro2020}. These two features are of recent introduction by our research group and are, in this work, for the first time tested in real-world conditions outside the laboratory.
Regarding the polarization encoder, the iPOGNAC is a novel design, based on a Sagnac polarization interferometer, that offers long-term stability and a low intrinsic Quantum Bit Error Rate (QBER). Furthermore, compared to other Sagnac-based polarization encoders~\cite{Agnesi2019,Li:19}, the iPOGNAC works with well defined polarization states meaning it does not require any calibration at the transmitter nor, for free-space channels, at the receiver.
On the other hand, regarding the synchronization between the transmitter and the receiver, the Qubit4Sync method is based on the exchange of qubits to share and determine the clock period and absolute time between the two parties. This method, somewhat similar to clock recovery schemes used in classical communications, allows for the synchronization of QKD setups without requiring additional hardware other than what is already necessary to prepare and measure the quantum states. By validating these novel and innovative techniques that reduce the complexity and improve the stability of QKD setups, the field-trial here described represents an important step towards mature, efficient and low-cost QKD systems.   

\begin{figure*}[t]

\includegraphics[width = \linewidth]{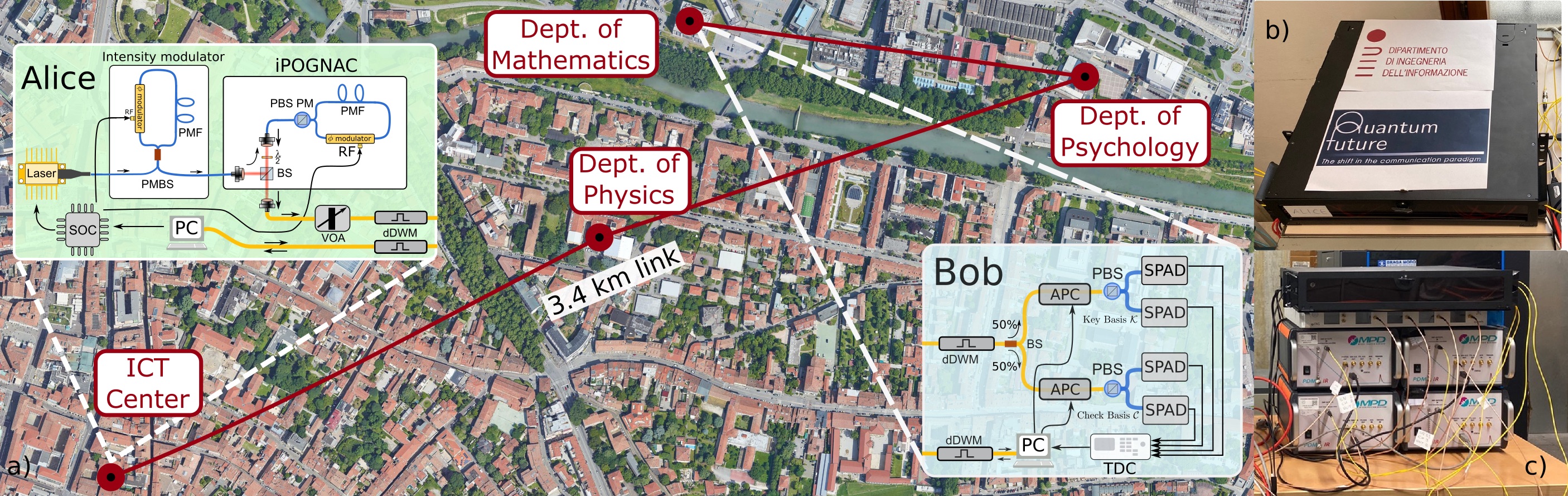}
\caption{\label{fig:setup} Schematic representation of the QKD field-trial performed in the metropolitan area of Padua, Italy. \textbf{a)} A map of the city center. The transmitter was placed at the ICT Center of the University of Padua while the receiver was located in the Department. of Mathematics. The transmitter and the receiver were connected by 3.4~km of deployed fibers. The insets show the experimental setups exploited during the field-trail.  Map Data from Google [\copyright2020 Google]. \textbf{b)} A photo of the transmitter fully contained in a 2U 19 inch rack enclosure. \textbf{c)} A photo of the receiver with its optical components fully contained in a 2U 19 inch rack enclosure.}
\end{figure*}

\section{Setup}

The experimental setup developed for the field-trial is illustrated in Fig.~\ref{fig:setup}. The transmitter (Alice), placed at the ICT Center of the University of Padua,  generates a train of optical pulses at 1550~nm, with 50~MHz repetition rate and $\sim 270$~ps (full-width-at-half-maximum) of temporal duration, using a gain-switched distributed feedback laser.  The intensity of each pulse is then set to one of the two possible levels required by the decoy-state method~\cite{Rusca2018} through an intensity modulator. In our case, the intensity modulator is based on a Sagnac interferometer~\cite{Roberts2018}  and is comprised of a 70:30 polarization maintaining beamsplitter (PMBS), a Lithium Niobate phase modulator and a 1m delay line of polarization maintaining fiber (PMF). This design allows for a temporally stable intensity modulation with a fixed ratio of $\sim 4.5$ between the mean photon number of the signal ($\mu$) and decoy ($\nu$) levels.  The polarization of the light pulses is then modulated via the iPOGNAC. It is important to note that up to this point the light pulses have traveled exclusively along the slow axis of PMFs, meaning that the light entering the iPOGNAC has a fixed and stable linear polarization state. The iPOGNAC is composed of both free-space and fiber segments. In particular, all components of the free-space segment are mounted on a steel mini-bench (Thorlabs FiberBench), which provides to the Alice setup compactness as well as mechanical stability. The light entering the iPOGNAC is directed to the free-space segment by a collimating lens. Here, the light pulses first encounter a half-wave plate which transforms the polarization state to a diagonal $\ket{D} = \left (\ket{H} + \ket{V} \right ) / \sqrt{2}$  polarization (where $\ket{H}$, $\ket{V}$ are the horizontal and vertical states respectively). Afterwards, the light pulses impinge a beamsplitter (BS) used to separate the input stream from the output stream. The light is therefore directed to a collimating lens that couples it into the fiber segment of the iPOGNAC. The fiber segment is composed of a PMF that leads to Polarizing Beamsplitter (PBS). This PBS marks the beginning of the Polarization Sagnac Interferometer which is composed of a Lithium Niobate phase modulator and a 1m PMF delay line. Inside the Sagnac loop, the two orthogonal polarizations travel in opposite directions and, due to the PMF delay line, traverse the phase modulator at different times. This allows us to individually address the phase ($\phi_e$ and $\phi_l$)  of each polarization component. The light is then recombined at the PBS, re-travels the PMF and is emitted back to the free-space channel by the collimating lens. The light then re-encounters the free-space BS which directs the light to the output port. The polarization state at the output is given by $\ket{\Psi_\mathrm{out}} = \left (\ket{H} + e^{i(\phi_l - \phi_e)} \ket{V} \right ) / \sqrt{2}$. Therefore, by applying a $\pi/2$ phase to $\phi_e$ or $\phi_l$ the states $\ket{L} = \left (\ket{H} + i \ket{V} \right ) / \sqrt{2}$ or $\ket{R} = \left (\ket{H} - i \ket{V} \right ) / \sqrt{2}$ can be generated. Instead, if no phase is applied, the state remains as $\ket{D}$. These three states are sufficient to implement the simplified version of BB84~\cite{Fung2006} with key-generating basis  $\mathcal{K}=\{\ket{L},\ket{R}\}$ and  check basis  $\mathcal C=\{\ket{D},\ket{A}=  \left (\ket{H} - \ket{V} \right ) / \sqrt{2} \}$.
At the output of the iPOGNAC, the light is coupled into a single mode fiber. To ensure the compatibility of the QKD transmitter with the classical infrastructure employed for the test, the output is filtered by a dense Wavelength Division Multiplexer (dWDM) centered around Channel 34 of the ITU\footnote{International Telecommunication Union-Telecommunication Standardization Bureau} grid, with a channel spacing of $100~\si{\giga\hertz}$. Finally, the light is attenuated at the single-photon level ($\mu = 0.487$, $\nu = 0.109$) by a variable optical attenuator (VOA) and injected into the quantum channel. The transmitter is controlled by a System-on-a-Chip (SoC) by Xilinx (Zynq-7000, FPGA+CPU), which generates the required control signals used to drive the intensity and polarization modulators, at a repetition rate of $50~\si{\mega \hertz}$. The random bits used for running the protocol are obtained from a source-device-independent quantum random number generator~\cite{Avesani2018}. 
The entire system is packaged in a compact and portable 2U 19 inch rack enclosure which can be seen in Fig.~\ref{fig:setup}b).
The metropolitan fibers deployed in the city center of Padua were used as a quantum channel, connecting the Department of Mathematics with the ICT Center of the University. The losses of the channel have been measured to be 9~dB, with a length of 3.4~km. 
At the receiver side (Bob), placed at the Department of Mathematics, a fiber-based optical setup measures the polarization state in the $\mathcal K$ or $\mathcal C$ basis, with $50\%$ probability. The received photons are initially filtered by a dWDM matched to the one installed in the transmitter. Then, the choice of the basis is done passively by a 50:50 beamsplitter, while two automatic polarization controllers, one after each output port of the beamsplitter (BS), are placed before a PBS to align the two measured basis.  The output of the optical measurement is detected by four PDM-IR InGas/InP single-photon avalanche detectors (SPADs) developed by Micro Photon Devices S.r.l, with 15\% quantum efficiency. The detectors are operated in free-running mode with a hold-off time of 20~$\mu$s. The time of arrival of the photon is recorded by a quTAG time to digital converter (TDC) developed by qutools GmbH.
The optical part of the receiver is packaged in a portable 2U 19 inch rack enclosure, similar to the one used for the transmitter. A photo of the detection system can be seen in Fig.~\ref{fig:setup}c.
Temporal synchronization and filtering is applied via software 
by exploiting the Qubit4Sync method~\cite{Calderaro2020}. In particular, we measured
 the temporal distribution of the detection events modulus 20~ns (the transmitter period) and eliminated those that lie outside a 750~ps window centered on the peak of this distribution.
This increases the SNR by eliminating $\frac{20~{\rm ns}-750~{\rm ps}}{20~{\rm ns}}=96.25\%$ of the events that are temporally uncorrelated with the signal and that are attributed to afterpulses, background light and dark counts.

\section{Field Trial Results}
\begin{figure}
\includegraphics[width = \columnwidth]{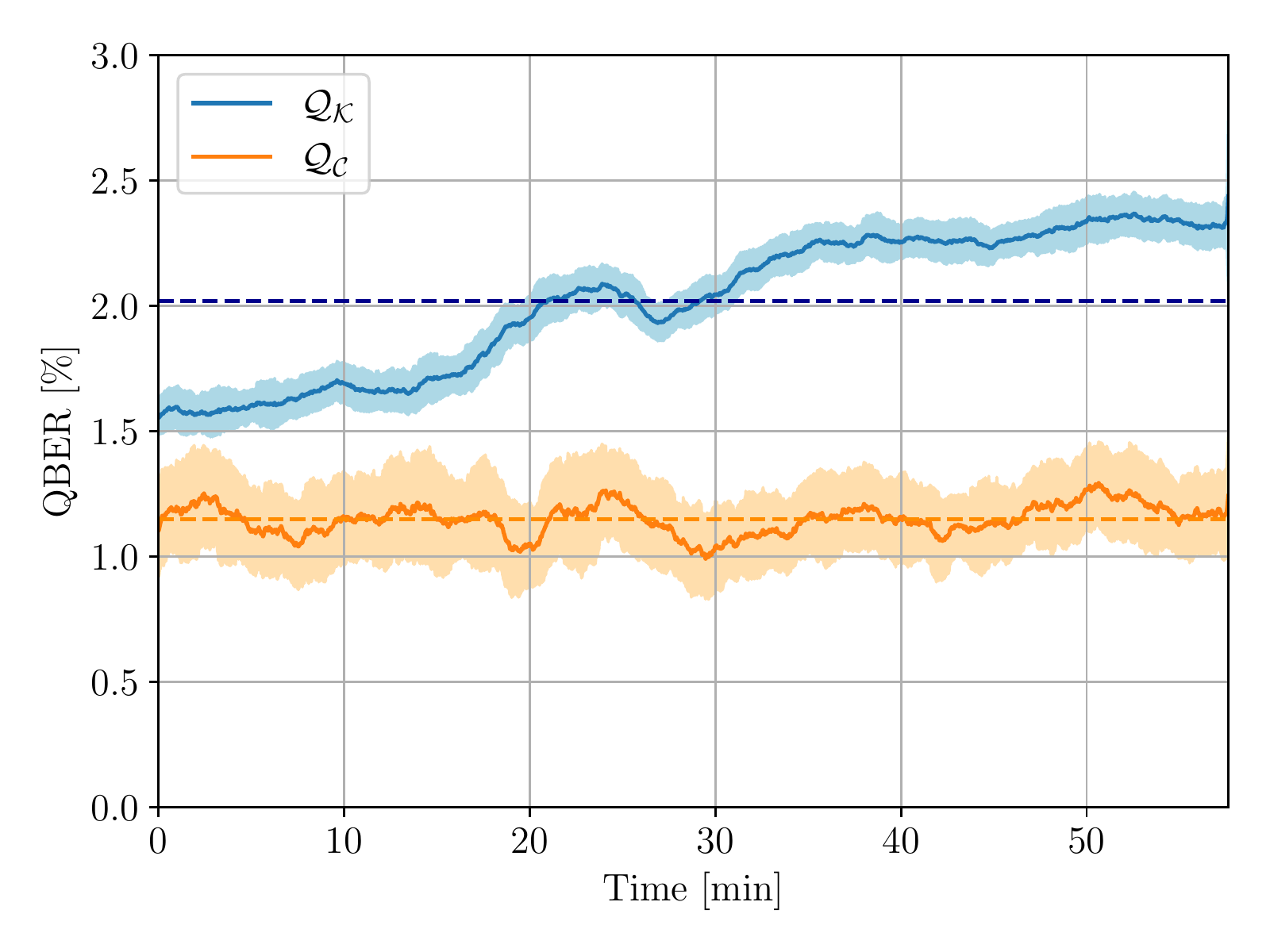}
\caption{\label{fig:qber} The QBERs $\mathcal{Q_\mathcal{K}}$ and $\mathcal{Q_\mathcal{C}}$ in the key and control bases respectively. They are measured every second and a rolling mean (60 s window) is plotted. Colored bands represent $\pm 1$ standard deviations, calculated on the rolling window. Dashed lines are the overall averages $2.0\%$ and $1.1\%$.}
\end{figure}

Typical duration for the  QKD system run is of one hour. As an example: in a run we accumulated a total of $2.7\cdot 10^8$ received qubits, which formed the raw keys.
In the post processing phase, we computed the quantum bit error rates (QBERs) $\mathcal{Q_\mathcal{K}}$ and $\mathcal{Q_\mathcal{C}}$ in the key and control bases respectively, shown in Fig. \ref{fig:qber}, whose overall averages were $2.0\%$ and $1.1\%$.
The analysis of the temporal distribution of the detection suggested that only a $0.5\%$ contribution to the QBER is due to noise (afterpulses, background light, dark counts), with the rest being attributable to alignment mismatches between the polarizations of the prepared states and the measurement bases.
We also noticed a slight increase in $\mathcal{Q_\mathcal{K}}$ during the hour-long acquisition and we attribute it to poorly compensated polarization drifts in the channel. 

We analysed the raw keys using a modified version of the AIT QKD R10 software suite by the AIT Austrian Institute of Technology GmbH~\cite{AIT}, tailored for this experiment. 
The classical communication happened on a fiber channel equal and parallel to the one used for the quantum transmission. 
The authentication of this channel was based on a pre-shared secret random sequence.
The post-processing algorithm produced approximately $4\cdot 10^7$ secret key bits, corresponding to a secret key rate (SKR) of 11.5 kbps.
This value is measured following the finite-size analysis of Rusca \textit{et al.}~\cite{Rusca2018}:
\begin{eqnarray}
    SKR &= [s_{\mathcal{K},0} + s_{\mathcal{K},1}(1 - h(\phi_\mathcal{K})) - \lambda_{\rm EC} \nonumber\\   &\quad\quad  -\lambda_{\rm conf} - 6 \log_2(19/\epsilon_{\rm sec})]/t  \ .
\end{eqnarray}
Here, $s_{\mathcal{K},0}$ and $s_{\mathcal{K},1}$ are the lower bounds on the number of vacuum and single-photon detections in the key basis, $\phi_\mathcal{K}$ is the upper bound on the phase error rate corresponding to single photon pulses, $h(\cdot)$ is the binary entropy, $\lambda_{\rm EC}$ and $\lambda_{\rm conf}$ are the number of bits published during the error correction and confirmation of correctness steps, $\epsilon_{\rm sec}$ is the security parameter associated to the secrecy analysis, which we set at $10^{-10}$, and finally $t$ is the duration of the quantum transmission phase. 
We did not include the duration of the classical post-processing because most of it can be parallelized with the quantum transmission phase, starting after a small delay.
Only the privacy-amplification phase must be done strictly after the end of the transmission of a long block of key, but its duration is negligible compared to that of the transmission, even when run on consumer-grade laptops.

\begin{figure}
\includegraphics[width = \columnwidth]{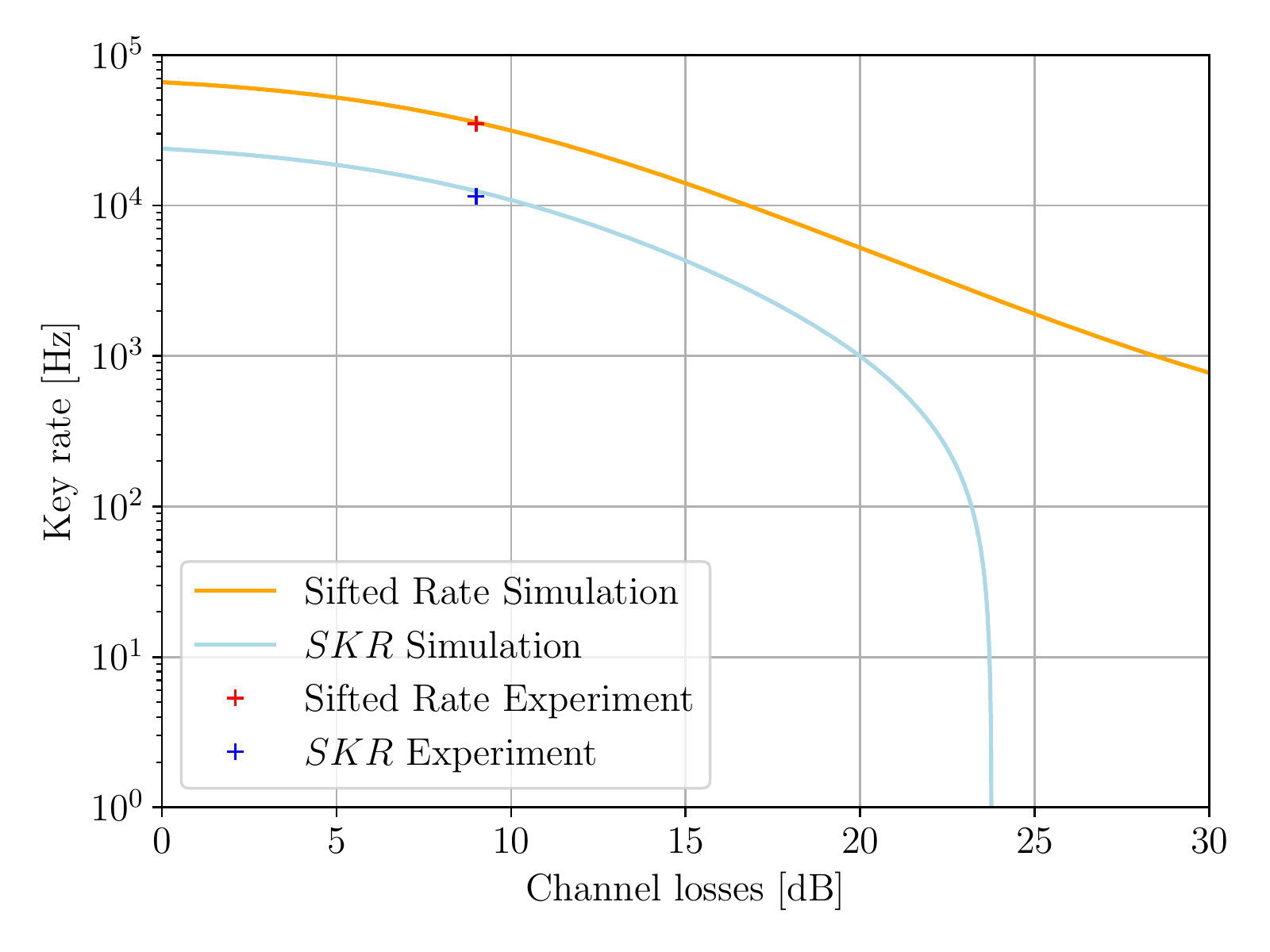}
\caption{\label{fig:simulation} Simulation of the performance of the system in terms of sifted and secret key rate, as a function of channel losses. All physical parameters are fixed and extracted from the experiment, with the exception of the channel losses. }
\end{figure}

Using parameters extracted from the experiment, we also simulated the performance of the system in different conditions of channel losses. 
As shown in Fig. \ref{fig:simulation}, we would be able to produce a secret key even with 23 dB of channel losses.

\section{Conclusions}
In this letter we have reported the development and in-field testing of a polarization-based QKD system implementing a simplified three-state and one-decoy BB84 QKD protocol~\cite{Grunenfelder2018}.
By exploiting a self-compensating design for both the intensity~\cite{Roberts2018} and polarization~\cite{Avesani:2020} modulators, the setup proved resilience to vibrations and temperature fluctuations. In fact, the setup did not require any re-calibration after transportation from the assembling location to the final locations and was able to generate quantum-secure cryptographic keys immediately after installation.
These design features combined with a dedicated control software, enabled the system to operate in a fully autonomous manner.
Furthermore, the system design also allowed for the entirety of the optical components of both the transmitter and the receiver to fit portable 2U 19 inch rack enclosures. This packaging, combined with the use of two dWDM channels for the quantum and classical communication, simplified the integration of our QKD system with the existing telecommunication infrastructure of the university.
The use of state-of-the-art InGaAs/InP SPADs combined with a highly effective temporal filtering allowed the receiver to operate the detectors in free-running mode. This eliminated the need of distributing a synchronization signal between Alice and Bob, enabling the use of the Qubit4Sync algorithm~\cite{Calderaro2020} for synchronization, without degrading the performances of the system. 
The results here presented attest that the described system is resource-efficient, valid and robust, and that it can be easily and rapidly installed in an existing telecommunication infrastructure, representing an important step towards the deployment in  telecommunication networks.

\begin{acknowledgments}
This work was supported by: European Union's Horizon 2020 research and innovation programme, project
{\it OpenQKD} (grant agreement No 857156); 
Agenzia Spaziale Italiana, project {\it Q-SecGroundSpace} (Accordo n. 2018-14-HH.0, CUP: E16J16001490001);
Ministero dell'Istruzione, dell'Universit\`a e della Ricerca under the initiative 
 ``Departments of Excellence'' (Law 232/2016); Fondazione Cassa
di Risparmio di Padova e Rovigo within the call ``Ricerca Scientifica di
Eccellenza 2018'',  project
{\it QUASAR}.
{M.~Z. acknowledges funding from the European Union's Horizon 2020 research and innovation programme under the Marie Sk\l{}odowska-Curie, grant agreement No 675662 {\it QCALL - Quantum Communications for ALL}}. 

We would like to sincerely thank Eng.~Giorgio Paolucci and Eng.~Fabian Ballarin of ASIT (Area Servizi informatici e telematici) - University of Padova and the Department of Mathematics of the University of Padova, the Director Prof.~Bruno Chiarellotto, Dr. Giulio Giusteri and the Network Team led by Eng.~Luca Righi for the logistic support during the realization of the field trial. 

We would like also to sincerely thank Dr.~Christoph Pacher and Dr.~Oliver Maurhart and the AIT Austrian Institute of Technology
GmbH for providing the foundation for the post-processing
software. CloudVeneto is acknowledged for the use of computing and storage facilities.

\end{acknowledgments}

\providecommand{\noopsort}[1]{}\providecommand{\singleletter}[1]{#1}%
\end{document}